\documentclass[useAMS,usenatbib]{mn2e}
\usepackage{times}
\usepackage{dcolumn}
\usepackage{graphicx}
\usepackage{aas_macros}
\usepackage{xspace}

\newcommand{\msun}{\mbox{${\rm M}_{\odot}$}\xspace}
\newcolumntype{d}{D{.}{.}{-1}}
\def\rev#1{#1}
\def\REV#1{#1}

\title[Reconstructing the evolution of white dwarf binaries]{
  Reconstructing the evolution of white dwarf binaries: further
  evidence for an alternative \rev{algorithm for the outcome of} the
  common-envelope phase in close binaries}

\author[G. Nelemans and C.A. Tout]  {G. Nelemans\thanks{Present
    address: Department of Astrophysics, Radboud University Nij\-me\-gen,
    The Netherlands
    E-mail: nelemans@astro.ru.nl} and C.A. Tout 
\\
  Institute of Astronomy, University of Cambridge, Madingley Road, Cambridge CB3 0HA, UK\\
 }

\defcitealias{npv+00}{NPVY}

\begin{document}

\date{Accepted . Received \today}

\pagerange{\pageref{firstpage}--\pageref{lastpage}} \pubyear{2004}

\maketitle

\label{firstpage}

\begin{abstract}
  We determine the possible masses and radii of the progenitors of
  white dwarfs in binaries from fits to detailed stellar evolution
  models and use these to reconstruct the mass-transfer phase in which
  the white dwarf was formed.  We confirm the earlier finding that in
  the first phase of mass transfer in the binary evolution leading to
  a close pair of white dwarfs, the standard common-envelope formalism
  \rev{(the $\alpha$-formalism)} equating the energy balance in the
  system \rev{(implicitly assuming angular momentum conservation)},
  does not work.  \rev{An algorithm equating} the angular momentum
  balance \rev{(implicitly assuming energy conservation)} can explain
  the observations.  This conclusion is now based on ten observed
  systems rather than three.  With the \rev{latter algorithm (the
    $\gamma$-algorithm)} the separation does not change much for
  approximately equal mass binaries. \rev{Assuming constant efficiency
    in the standard $\alpha$-formalism and a constant value of
    $\gamma$,} we investigate the effect of both \rev{methods} on the
  change in separation in general and conclude that when there is
  observational evidence for strong shrinkage of the orbit, the
  \rev{$\gamma$-algorithm} also leads to this.  We \rev{then} extend
  our analysis to \emph{all} close binaries with at least one white
  dwarf component and reconstruct the mass transfer phases that lead
  to these binaries. \rev{In this way we find all possible values of
    the efficiency of the standard $\alpha$-formalism and of $\gamma$
    that can explain the observed binaries for different progenitor
    and companion masses. We find that \emph{all} observations can be
    explained with a single value of $\gamma$, making the
    $\gamma$-algorithm a useful tool to predict the outcome of
    common-envelope evolution.} 
%We discuss the physical basis of the
%  two \rev{methods} and show that \rev{\emph{if} after the onset of
%    Roche-lobe overflow -- but before the physical engulfment of the
%    companion in the envelope -- there can be a relatively slow early
%    phase where angular momentum dominates the evolution, a
%    combination of Jeans mass loss and the standard
%    $\alpha$-formalism} might be physically plausible and gives
%  results that are in agreement with all the observations.  
We discuss
  the consequences of our findings for different binary populations in
  the Galaxy, including massive binaries, for which the reconstruction
  method cannot be used.
\end{abstract}


\section{Introduction}

Now that double white dwarfs are discovered regularly
\citep[e.g.][]{mar95,mar00,ncd+01} it has become more and more clear
that most of them have a mass ratio close to unity
\citep*[e.g.][]{mm99,mmm02}. This is contrary to what is expected from
standard population synthesis calculations
\citep*[e.g.][]{ity97,han98}. A possible resolution of this issue was
investigated by \citet{nvy+00,nyp+00}. In the first paper the observed
masses of three double white dwarfs and the well known core-mass --
radius relation were used to reconstruct the evolution of the binary
back to two main-sequence stars. It followed that the first phase of
mass transfer could not be described by the standard common-envelope
formalism \rev{(based explicitly on energy balance, assuming angular
  momentum conservation implicitly)}, nor by stable Roche-lobe
overflow. \rev{Recent calculations using a detailed stellar evolution
  code have confirmed this conclusion (Van der Sluys et al., in
  preparation). Stable Roche-lobe overflow leads to final double white
  dwarfs with a mass ratio larger than one \citep[e.g.][]{ity97,han98}
  and the observed masses can only be reached by stars with initial
  masses between about 2.3 and 3.5 \msun that fill their Roche-lobes
  within a very small initial separation interval (in order to start
  mass transfer in the Hertzsprung gap). For standard population
  synthesis assumptions \citep[e.g.][]{nyp03} this interval only
  accounts for 0.3 per cent of objects forming white dwarfs, so is
  inconsistent with the observation that about 10 per cent of white
  dwarfs are close pairs \citep{mm99}.}  Instead \citet{nvy+00}
proposed an \rev{empirical} algorithm based \rev{explicitly} on
angular momentum balance \rev{(implicitly assuming energy
  conservation) with a single free parameter and concluded that all
  the observed systems could be explained with the same value of the
  free parameter.}  The second paper showed that \rev{using this
  algorithm} a satisfactory model for the Galactic population of
double white dwarfs can be obtained.

Since then, quite a few more double white dwarfs have been discovered.
In particular the SPY project \citep{ncd+01}, a large survey on the
ESO Very Large Telescope, to measure radial velocity variations of
some thousand white dwarfs in order to detect duplicity has, and will,
enlarge the known double white dwarf sample. We therefore repeat the
analysis of \citet{nvy+00}, including the new discoveries
(Section~\ref{reconstruct}). Furthermore, we study the difference
between the \rev{standard and alternative method} in some detail
(Section~\ref{formalisms}).  We then extend the analysis to \emph{all}
binaries with at least one white dwarf component
(Section~\ref{WD_binaries}) and sdB binaries
(Section~\ref{sdB_binaries}) \rev{in order to determine what the free
  parameter in the alternative method must be to explain the
  observations.}  We then continue with a 
%\rev{brief} discussion of
%our findings in terms of the physics of the mass transfer, a
%comparison with model calculations (Section~\ref{theory}) and a
discussion of the consequences of the our results for the different
binary populations (Section~\ref{consequences}) and round off with our
conclusions.

%\begin{itemize}
%\item Origin of common envelope idea
%\item Types of binaries affected
%\item learning more: theory vs observations (population synthesis)
%\item mention reconstruction double white dwarfs
%\item structure of the paper
%\end{itemize}

\section{Reconstruction of the evolution of double white dwarf binaries}\label{reconstruct}

\begin{figure*}
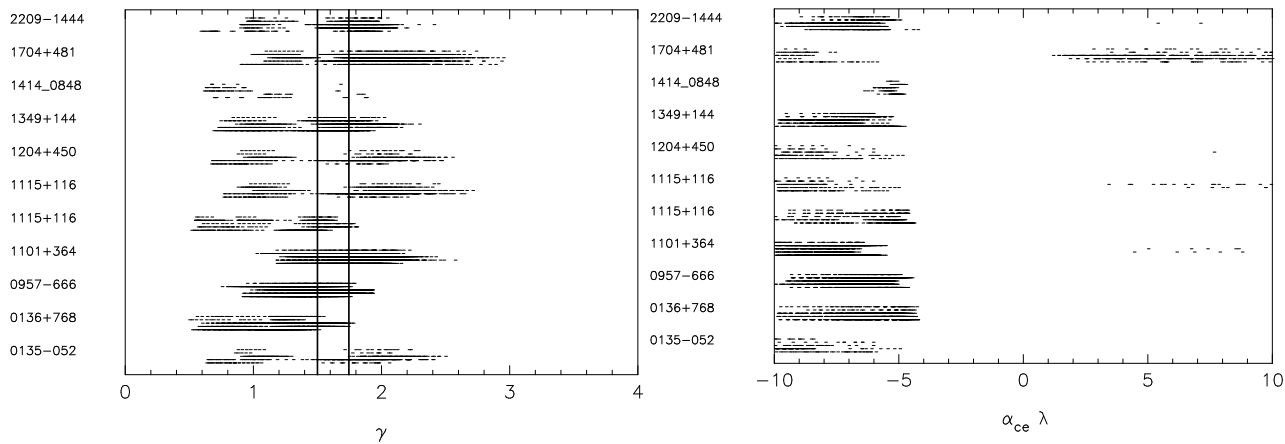

\resizebox{\columnwidth}{!}{\includegraphics[angle=-90]{gamma_wdwdfirst_1211.ps}}
\resizebox{\columnwidth}{!}{\includegraphics[angle=-90]{al_wdwdfirst_1211.ps}}
\caption[]{Left: Reconstructed $\gamma$ values for the first phase of mass
  transfer in the formation of double white dwarfs. Right:
  reconstructed $\alpha \lambda$ values for the same. The horizontal
  lines are made up of small dashes representing the reconstructed
  values of $\gamma$ and $\alpha \lambda$ for different values of the
  mass of the progenitor of the white dwarf and the companion. The
  different lines for each object represent different values of the
  white dwarf mass (within 0.05 \msun of the value in
  Table~\ref{tab:dwd}). }
\label{fig:gamma_al_wdwd_first}
\end{figure*}

We start with a short revision of the method used in \citet{nvy+00}.
The fact that observed white dwarfs in binaries were the cores of the
giant stars from which they descend makes it possible to reconstruct
the properties, in particular the radii, of these giants. On the
assumption that the observed white dwarf mass is close to the mass of
the core of the giant at the onset of mass transfer (i.e. that the
mass-transfer proceeds on a short time-scale \rev{compared to the
nuclear evolution time}), the exact evolutionary phase (and thus mass
and radius) of the giant at that instant can be obtained for each
possible initial progenitor mass from single star evolution models.
For a Roche-lobe filling giant its radius, together with its mass and
that of the companion, determine what the orbital separation at the
onset of mass transfer. By comparing this with the orbital separation
after the mass transfer, the effect of the common-envelope phase on
the orbit can be reconstructed.

In \citet{nvy+00} only double helium white dwarfs were considered and
a simple core-mass -- radius relation for giants with degenerate
helium cores was used to reconstruct the properties of the giants.
Here we take a more general approach and use fits to detailed stellar
evolution calculations to find all the possible giant stars that have
a core with a mass equal to the observed white dwarf mass. For this we
use the \citet*{hpt00} fits which enable us to use all observed double
white dwarfs, independent of them being helium or \rev{(low-mass)}
carbon-oxygen white dwarfs.

Our exact procedure is as follows. For an observed white dwarf mass
$M_{\rm WD}$ we use the \citet{hpt00} equations to calculate the
masses $M_{\rm giant}$ and radii $R_{\rm giant}$ of all the giants
which have exactly such a core mass.  We do this for initial masses of
1, 1.1, 1.2, ...\msun up to the mass for which the initial core mass,
at the end of the main sequence, is larger than the observed white
dwarf mass. While evolving the stars we keep track of the maximum
radius the star has reached previously so that only giants that
actually can fill their Roche lobe are selected.  Finally, we only
consider stars if they have passed through the Hertzsprung gap and
have developed convective envelopes. Radiative stars in the
Hertzsprung gap \REV{can avoid a common-envelope phase so that} our
assumption of mass transfer on a short time-scale compared to the
evolutionary time-scale is not appropriate.

For each of the possible masses $m$ for the companion (see below) we
use the size of the Roche lobe $R_L$ in units of the separation $a$,
$r_L = R_L/a$, as given by \citet{egg83} to determine the separation
at the onset of the mass transfer assuming $R_{\rm giant} = R_L$. The
range of companion masses considered is determined by the
observations.  If the mass of the companion is known that mass is used
but for unseen companions in double white dwarf systems we use the
extremes of 0.2 and 1.4 \msun as in \citet{nvy+00}.

\begin{table}
\caption[]{Properties of the observed double white dwarfs}
\label{tab:dwd}
\begin{center}
\begin{tabular}{ldddr} \hline
Object (WD/HE)   & \multicolumn{1}{c}{$P$} &
\multicolumn{1}{c}{$M_{\rm WD, 2}$} & \multicolumn{1}{c}{$M_{\rm WD, 1}$} &  Ref \\ 
 &  \multicolumn{1}{c}{(d)} &
\multicolumn{1}{c}{(\msun)} & \multicolumn{1}{c}{(\msun)} & \\ 
\hline
0135$-$052 &  1.56 &  0.47 & 0.52 & 1,2  \\
0136$+$768 &  1.41 &  0.47 & 0.37 & 8,13 \\
0957$-$666 &  0.06 &  0.37 & 0.32 & 3,7  \\
1022$+$050 &  1.16 &  0.35 &      & 8    \\
1101$+$364 &  0.15 &  0.29 & 0.35 & 4,13 \\
1115$+$116 & 30.09 &  0.52 & 0.43 & 12   \\
1202$+$608 &  1.49 &   0.4 &      & 6    \\
1204$+$450 &  1.60 &  0.46 & 0.52 & 8,13 \\
1241$-$010 &  3.35 &  0.31 &      & 5    \\
1317$+$453 &  4.87 &  0.33 &      & 5    \\
1349$+$144 &  2.12 &  0.44 & 0.44 & 14   \\
1414$-$0848&  0.518&  0.71 & 0.55 & 11,15\\
1428$+$373 &  1.143&  0.33 &      & 9    \\
1704$+$481 &  0.14 &  0.39 & 0.56 & 10   \\
1713$+$332 &  1.12 &  0.35 &      & 5    \\
1824$+$040 &  6.27 &  0.39 &      & 8    \\
%2020$-$425 &  0.30 &  0.52 & 0.93 & 17   \\ 
2032$+$188 &  5.084&  0.36 &      & 8    \\
2209$-$1444&  0.28 &  0.58 & 0.58 & 16   \\
2331$+$290 &  0.17 &  0.39 &      & 5    \\ \hline
\end{tabular}
\end{center}

References: (1) \citet*{slo88}; (2) \citet{bwf+89}; (3) \citet{bgr+90};
(4) \citet{mar95}; (5) \citet*{mdd95}; (6) \citet{hst+95}; (7)
\citet*{mmb97}; (8) \citet{mm99}; (9) \citet{mar00} and P. Maxted,
private communication; (10)
\citet{mmm+00}; (11) \citet{nkn02}; (12) \citet{mbm+02}; (13)
\citet*{mmm02}; (14) \citet{knh+02}; (15) \citet{ndh+02}; (16)
\citet{knn03}; 
%(17) Napiwotzki, priv. comm
\end{table}

In Table~\ref{tab:dwd} we list the properties of the observed double
white dwarfs. It includes both updates and additions to table~1 of
\citet{nvy+00}. There are now 10 binaries in which the masses of both
components are known. For these we can use our reconstruction method
twice, first for the last phase of mass transfer in which the white
dwarf with mass $M_{\rm WD, 2}$ is formed and the companion star was a
white dwarf of mass $M_{\rm WD, 1}$. This gives the separation before
the second phase of mass transfer and the mass of the giant that
formed white dwarf~2. We then calculate the separation after the
\emph{first} phase of mass transfer by assuming the separation only
changed owing to mass loss in a wind from the progenitor of white
dwarf~2. Finally we use the initial mass of the progenitor of white
dwarf~2 and the mass of white dwarf~1 ($M_{\rm WD, 1}$) to calculate
the change in separation in the first phase of mass transfer. The only
extra constraint we have to put in is that we require the progenitor
of white dwarf~1 to be more massive than the reconstructed progenitor
of white dwarf~2.

We now discuss the results for the first phase of mass transfer
because that is the phase that was found to be inconsistent with the
standard common-envelope formalism, proposed by \citet{pac76} to
explain the existence of short-period binaries with white dwarf
components and cataclysmic variables. It is generally assumed that the
outcome of the common-envelope phase is determined by the energy
balance, implicitly assuming angular momentum conservation. I.e. that
the orbital energy of the binary is used to expel the envelope of the
giant with some efficiency $\alpha$ \citep[e.g.][]{web84}
\begin{equation}\label{eq:Energy}
\frac{G M_{\rm g} M_{\rm e}}{\lambda R_{\rm g}} = \alpha
\left(\frac{G M_{\rm c} m}{2 a_{\rm f}} - \frac{G M_{\rm g} m}{2 a_{\rm i}}\right),
\end{equation}
where subscripts g, e and c are for giant, envelope and core
respectively and we assume the companion mass does not change during
the common-envelope phase. The structural parameter $\lambda$ is
normally taken as a constant \citep*[e.g. $\lambda = 0.5$][]{khp87},
or as \citet{nvy+00} the $\lambda$ factor is incorporated in the
uncertain efficiency factor to give one free parameter $\alpha
\lambda$ and this is what we do here. \rev{We will refer to this method
as the standard $\alpha$-formalism}.

The algorithm based \rev{explicitly} on the equation for angular momentum
balance (implicitly assuming energy conservation, but not necessarily
only for orbital and binding energy) proposed by \citet{nvy+00} is
described by
\begin{equation}\label{eq:AM}
\frac{\Delta J}{J} = \gamma \frac{\Delta M_{\rm total}}{M_{\rm total}}
  = \gamma \frac{M_{\rm e}}{M_{\rm g} + m}.
\end{equation}
\rev{In the remainder of the paper we refer to this method as the
$\gamma$-algorithm}.

For each double white dwarf for which the masses of both white dwarfs
are known we can calculate the range of possible masses of the
secondary from $M_{\rm WD,2}$ and separations \emph{after} the first
phase of mass transfer and the possible masses and radii of the
primary from $M_{\rm WD, 1}$ and thence the separation \emph{at the
  onset} of the first phase of mass transfer. That means that all
terms in equation~(\ref{eq:Energy}) except $\alpha \lambda$ and all terms
in equation~(\ref{eq:AM}) except $\gamma$ are known. For the calculation of
the total angular momentum we include the angular momentum of the
giant, assuming all the angular momentum resides in the envelope,
which we approximate as an $n = 3/2$ polytrope. In
Fig.~\ref{fig:gamma_al_wdwd_first} we show, for each of the observed
double white dwarfs, the possible values of $\alpha \lambda$ and
$\gamma$ that we find in this way. Each possible combination of
progenitor and companion mass is shown as a small dash, forming
horizontal lines. The different lines for each object are for
different values of the white dwarf mass to account for measurement
errors (which we take as $\pm$0.05 \msun). WD1115+116 is shown twice
because it is not clear from the observations which of the two white
dwarfs is white dwarf~1 and which is white dwarf~2.

We confirm the findings of \citet{nvy+00} that the first phase of mass
transfer in the evolution leading to the observed double white dwarfs
cannot generally be described by the \rev{standard $\alpha$-formalism}
because the reconstructed values of $\alpha \lambda$ are negative.
The only exception is WD1704+481 which does have a mass ratio in the
range expected from evolution governed by the standard
$\alpha$-formalism.

\begin{figure}
\resizebox{\columnwidth}{!}{\includegraphics[angle=-90]{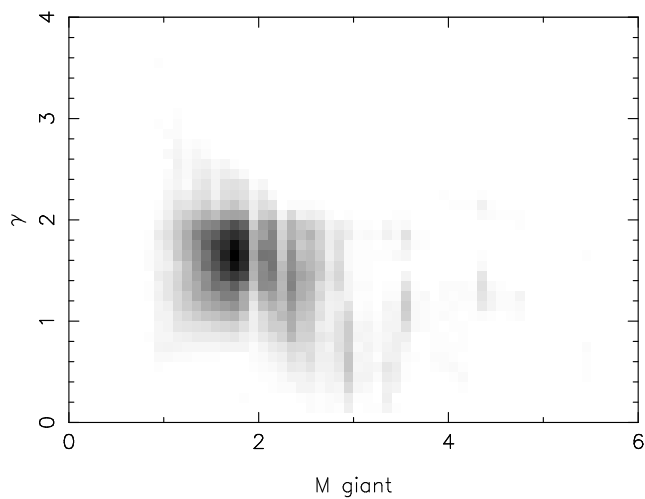}}
%\\
%\resizebox{\columnwidth}{!}{\includegraphics[angle=-90]{hist_gam_wdwdfirst.ps}}
\caption[]{Reconstructed $\gamma$ values versus the mass of the
  giant for the first phase of mass transfer in the formation of
  double white dwarfs.}
\label{fig:Mgiant_gamma_wdwdfirst}
\end{figure}

\begin{figure*}
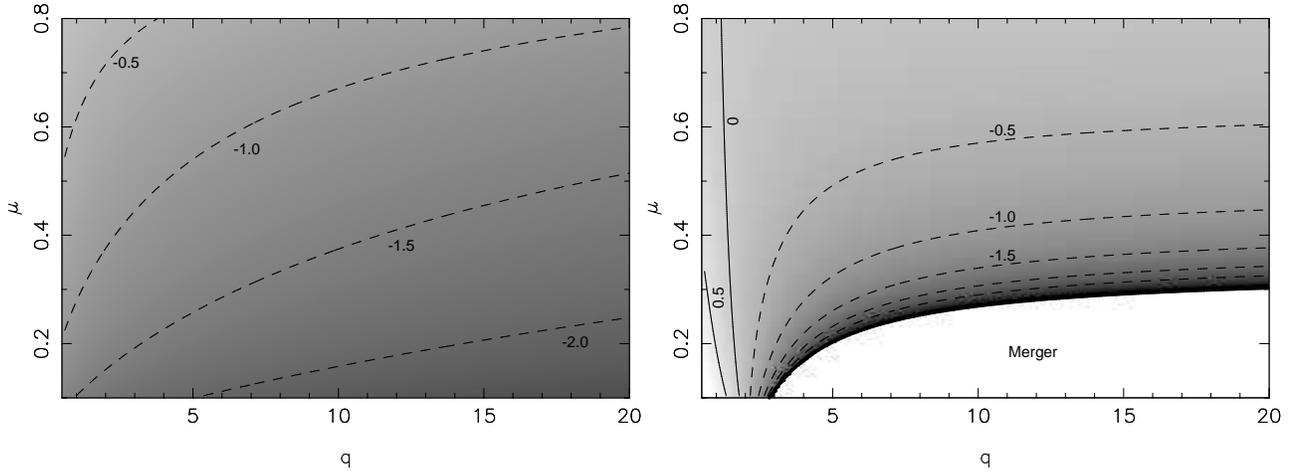

\resizebox{\columnwidth}{!}{\includegraphics[angle=-90]{al.ps}}
\resizebox{\columnwidth}{!}{\includegraphics[angle=-90]{gam.ps}}
\caption[]{Relative change in separation for dynamical mass transfer
  described by the \rev{standard $\alpha$-formalism} (left panel) and the
  \rev{$\gamma$-algorithm} (right panel) as function of mass ratio $q$ and
  core mass fraction $\mu$. The logarithm of $(a_{\rm i}/a_{\rm f})$
  is shown as the grey scale and the contours. }
\label{fig:ai_af}
\end{figure*}

As to the values of $\gamma$, we also recover the results of
\citet{nvy+00} with typical values around 1.5 with a large spread.
However for a value of $\gamma$ between 1.5 and 1.75 we can find
simultaneous solutions for all objects.
% except WD2020-425 in case
%the massive white dwarf is white dwarf 2, for which there is no
%solution using the procedure explained above, as the progenitor of the
%0.93 \msun white dwarf, which should be the secondary is always more
%massive than any possible progenitor of the 0.53 \msun white dwarf. If
%the more massive white dwarf is indeed the younger, this system could
%have formed through a stable first phase of mass transfer in which the
%secondary accreted a substantial amount of mass.

To asses the likelihood of the solutions found with $\gamma$ around
1.5 we plot the mass of the giant versus the reconstructed $\gamma$
value in Fig.~\ref{fig:Mgiant_gamma_wdwdfirst}. Typical giant masses
are between 1.5 and 2 \msun, just as one would expected for the more
massive components in binaries that eventually form double white
dwarfs, i.e. in which both stars evolve off the main sequence within
the age of the Galaxy.

%\begin{itemize}
%\item Giant core-mass -- radius relation, white dwarf mass --
%  progenitor radius relation
%\item In binary: white dwarf mass -- pre-mass-transfer separation
%  relation
%\item Assumptions: no mass accreted, mass transfer dynamic
%\item Equations
%\end{itemize}

\begin{figure*}
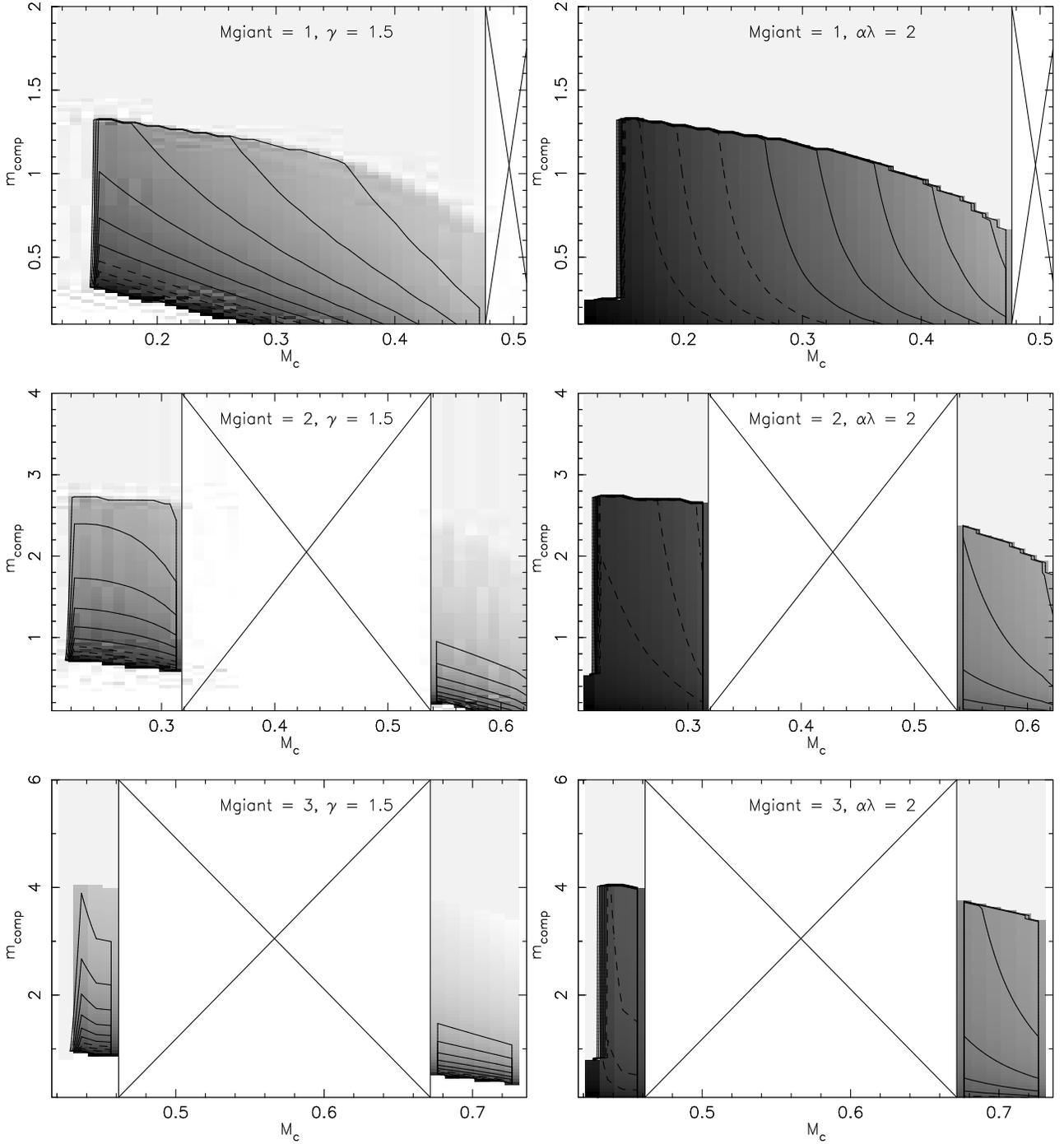

\resizebox{\columnwidth}{!}{\includegraphics[angle=-90]{Mc_mP_1WD_gam.ps}}
\resizebox{\columnwidth}{!}{\includegraphics[angle=-90]{Mc_mP_1WD_al.ps}}
\\
\resizebox{\columnwidth}{!}{\includegraphics[angle=-90]{Mc_mP_2WD_gam.ps}}
\resizebox{\columnwidth}{!}{\includegraphics[angle=-90]{Mc_mP_2WD_al.ps}}
\\
\resizebox{\columnwidth}{!}{\includegraphics[angle=-90]{Mc_mP_3WD_gam.ps}}
\resizebox{\columnwidth}{!}{\includegraphics[angle=-90]{Mc_mP_3WD_al.ps}}
\caption[]{Final periods as function of core mass and companion mass
  with the \rev{$\gamma$-algorithm} (left) and the \rev{standard $\alpha$-formalism} (right)
  for giants of 1 \msun (top), 2 \msun (middle) and 3 \msun (bottom).
  The darkest shades represent periods of 0.01 d and the lightest
  periods of 1000 d. Dashed contours are for constant log $(P/d)$ =
  -2, -1.5 -0.5 and the solid contours are for log $(P/d)$ = 0, 0.5,
  ..., 2.5. The even light gray area shows parts of parameter space
  for which stable mass transfer is expected. The gap in the middle
  occurs because the core mass grows during stages (core helium
  burning) when the star has a smaller radius than it had before and
  Roche-lobe overflow cannot take place. The white area below the
  shaded areas denotes combinations for which the systems merge when
  using the \rev{$\gamma$-algorithm} because all the angular momentum is
  lost from the system.}
\label{fig:Mc_m}
\end{figure*}

\section{Comparison of the \rev{standard $\alpha$-formalism} and the \rev{$\gamma$-algorithm}}\label{formalisms}

\citet{nvy+00} proposed that the first phase of mass transfer in the
evolution to a close double white dwarf was special in the sense that
it is most likely a phase of dynamical mass transfer but in a binary
with mass ratio not too far from unity. In such a binary the angular
momentum of the orbit is so large that the envelope of the giant can
be spun up easily. This removes the drag forces that might drive any
loss of orbital energy. We \REV{will discuss} the question of the physical
interpretation and applicability of the \rev{$\gamma$-algorithm} \REV{in
a forthcoming paper but here} we consider the effect of both \rev{methods}
described above on the change in orbital separation for a wide range
of giant and core masses and mass ratios.

The change in separation of the binary for the \rev{standard $\alpha$-formalism} is
\begin{equation}\label{eq:ai_af_Energy}
\left(\frac{a_{\rm f}}{a_{\rm i}}\right)_{\alpha} = \frac{M_{\rm
    c}}{M_{\rm g}} \left( 1 + \frac{2
      M_{\rm e}}{\alpha \lambda r_{\rm L} m} \right)^{-1},
\end{equation}
where we used $R_{\rm g} = R_{\rm L, giant} = r_{\rm L} a_{\rm i}$
and we have again assumed that none of the envelope is accreted by the
companion. For the \rev{$\gamma$-algorithm} this ratio  is
\begin{equation}\label{eq:ai_af_AM}
\left(\frac{a_{\rm f}}{a_{\rm i}}\right)_{\gamma} = \left(\frac{M_{\rm
      g}m}{M_{\rm c}
      m}\right)^{2} \left(\frac{M_{\rm c} + m}{M_{\rm g} + m}\right) \left(1 -
    \gamma \frac{M_{\rm e}}{M_{\rm g} + m}\right)^{2}.
\end{equation}
%where $\Delta M = M_{\rm e}$ again.

%\begin{figure}
%\resizebox{\columnwidth}{!}{\includegraphics[angle=-90]{al.ps}}
%\caption[]{Relative change in separation for dynamical mass transfer
%  described by the standard formalism as function of mass ratio $q$ and
%  core mass fraction $\mu$. The logarithm of $(a_{\rm i}/a_{\rm
%    f})_{\rm Energy}$ is shown as the grey scale and the contours.}
%\label{fig:ai_af_al}
%\end{figure}

%\begin{figure}
%\resizebox{\columnwidth}{!}{\includegraphics[angle=-90]{gam.ps}}
%\caption[]{Relative change in separation for dynamical mass transfer
%  described by the angular momentum formalism as function of mass
%  ratio $q$ and core mass fraction $\mu$. The logarithm of $(a_{\rm
%    i}/a_{\rm f})_{\rm AM}$ is shown as the grey scale and the contours.}
%\label{fig:ai_af_gam}
%\end{figure}

Only ratios of the masses of the different components (giant,
companion, core, and envelope) enter these equations, so that the
relative change in the orbital separation does not depend on the total
mass in the system but only on the mass ratios of the different
components.  There are only three independent masses (companion mass,
giant mass and either core mass or giant envelope mass). These are
characterised by only two ratios and the ratios $q = M_{\rm g}/m$ and
$\mu = M_{\rm core}/M_{\rm g}$ or alternatively $\Delta = M_{\rm
  e}/M_{\rm giant} = 1 - \mu$ conveniently simplify
equations~(\ref{eq:ai_af_Energy},\ref{eq:ai_af_AM}):
\begin{equation}
\left(\frac{a_{\rm f}}{a_{\rm i}}\right)_{\alpha} = (1 - \Delta) \left(1 +
    \frac{2 \Delta q}{\alpha \lambda r_{\rm L}(q)}\right)^{-1}
\end{equation}
and
\begin{equation}
\left(\frac{a_{\rm f}}{a_{\rm i}}\right)_{\gamma} = \left(\frac{1 - \gamma
      \frac{\Delta q}{1 + q}}{1 - \Delta}\right)^{2} \left(1 -
      \frac{\Delta q}{1 + q}\right).
\end{equation}

With these equations we calculate the change in separation for both
\rev{methods} as functions of $q$ and $\mu$, using $\alpha\lambda = 2$
and $\gamma = 1.5$. The results are shown in Fig.~\ref{fig:ai_af}. The
grey scale denotes the log of $a_f/a_i$, with darker shades a greater
shrinkage of the orbit. Contours of constant $\log a_{\rm f}/a_{\rm
  i}$ of 0.5, 0, $-$0.5 etc are also shown in the figures.
Fig.~\ref{fig:ai_af} shows that for any combination of $q$ and $\mu$
the \rev{standard $\alpha$-formalism} gives a strong shrinkage of the orbit
while with the \rev{$\gamma$-algorithm} there is a wide range from expansion
to very extreme shrinkage (or even guaranteed merger if all the
angular momentum is lost).

The reconstruction of double white dwarfs discussed in
Section~\ref{reconstruct} finds evidence for a strong reduction in the
last but not in the first phase of mass transfer. For typical
progenitors of double white dwarfs, with $\mu$ between 0.2 and 0.5 and
$q$ between 1 and 2 in the first and between 2 and 6 for the second
phase of mass transfer the \rev{standard $\alpha$-formalism} would give strong
shrinkage of the orbit in both cases.  In contrast, the
\rev{$\gamma$-algorithm} gives widening or very mild shrinkage in the first,
\emph{but strong shrinkage in the second phase of mass transfer} and
thus might explain both phases in the evolution to double white
dwarfs.

\begin{figure*}
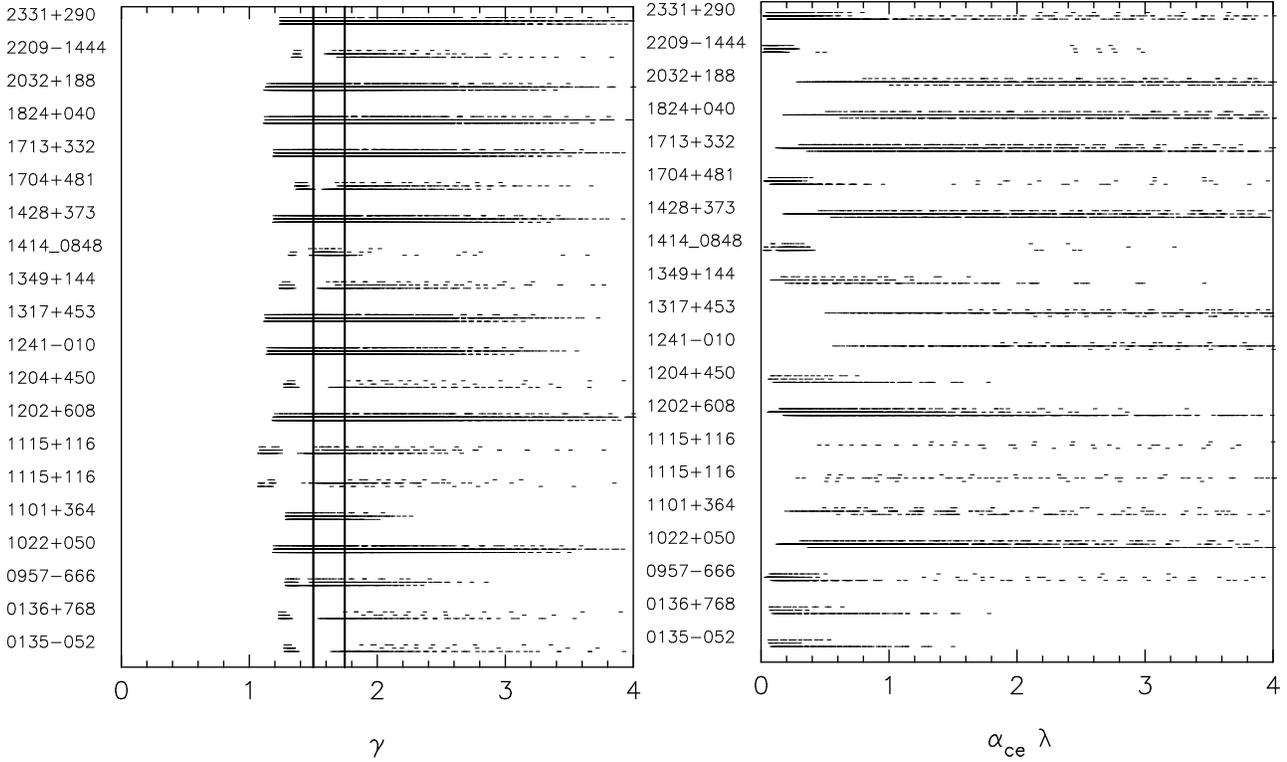

\resizebox{\columnwidth}{!}{\includegraphics[]{gamma_wdwdlast_1211.ps}}
\resizebox{\columnwidth}{!}{\includegraphics[]{al_wdwdlast_1211.ps}}
\caption[]{Left: Reconstructed $\gamma$ values for the last phase of mass
  transfer in the formation of double white dwarfs (see
  Table~\ref{tab:dwd}). Right: reconstructed $\alpha \lambda$ values
  for the same.}
\label{fig:gamma_al_wdwd_last}
\end{figure*}

We explore the difference between the two \rev{methods} further in
Fig.~\ref{fig:Mc_m} where, for initial masses of 1, 2, and 3 \msun, we
plot a grey scale of the period at the end of the common-envelope
phase for both \rev{methods} as function of the core mass and companion
mass. We again use $\gamma = 1.5$ and $\alpha \lambda = 2$. It can be
seen that for the \rev{standard $\alpha$-formalism} the final periods are below 10 d
except for the most massive cores, while for the \rev{$\gamma$-algorithm} it
depends strongly on the mass of the companion. Indeed, for relatively
high core masses and companion masses, very large final periods, above
1000 d, can be induced. This is interesting in the light of the
existence of symbiotic binaries, barium and S-stars with periods in
that range. The alternative for the formation of these binaries is
that they avoided a common-envelope phase. That is they have stable
mass transfer or avoid mass transfer at all and would be expected to
have even longer orbital periods. We will come back to these binaries
in Section~\ref{consequences}.

The observational requirement for strong orbital shrinkage has always
been for rather extreme-mass-ratio systems such as cataclysmic
variables and low-mass X-ray binaries. So the fact that the
\rev{$\gamma$-algorithm} actually produces a strong shrinkage at large mass
ratios makes it useful to consider the \rev{$\gamma$-algorithm} in more
extreme mass ratio common-envelope phases, such as the last phase of
mass transfer leading to a close double white dwarf and mass transfer
in binaries leading to a close binary with a white dwarf and a
main-sequence star, many of which are observed.  We can use the same
procedure we used to reconstruct the first phase of mass transfer in
double white dwarfs to reconstruct any of these.

\section{White dwarf binaries}\label{WD_binaries}

There are two classes of objects for which the standard
$\alpha$-formalism has been successfully used to explain their
properties. These are the last phase of mass transfer leading to the
formation of a close double white dwarf \citep[e.g.][]{nyp+00} and the
formation of close white dwarf -- main-sequence binaries. The latter
are expected to be the precursors of cataclysmic variables
\citep[e.g.][]{kr93}.  We discuss these binaries here in turn,
comparing again the \rev{standard $\alpha$-formalism} and the
\rev{$\gamma$-algorithm in order to asses how well they do in
  predicting the outcome of the common-envelope phase}.

There are two complicating factors which we have to take into account,
the first of which is tidal interaction. If the spin angular momentum
of one of the components in a binary exceeds one third of the orbital
angular momentum, the tidal interaction is unstable
\citep[see][]{hut80}.  \citet{ss74} showed that for mass ratio's
larger than about 6, the tidal instability sets in before the giant
fills its Roche lobe. Some of the binaries we shall discuss must have
had quite extreme mass ratios at the onset of the mass transfer
because the companions are either low-mass white dwarfs or low-mass
main-sequence stars. We therefore build in a check for tidal stability
in the reconstruction process. When a progenitor system is found to be
tidally unstable for our assumption that the onset of the mass
transfer is caused by Roche-lobe overflow, we relax this assumption
and instead assume the mass transfer was caused by the tidal
instability and we calculate the initial separation at which the
instability sets in at exactly the right core mass. The last aspect of
the new procedure is a check whether the initial separation is small
enough that the companion will actually keep the giant in co-rotation
with the orbit, because otherwise the tidal instability will not set
in at all and mass transfer is avoided. We use the maximum separation
as given in \citet{nt98}, based on \citet{zah77}.

The second is the question whether the current orbital period of the
observed systems is a good estimate of the post-mass-transfer period.
In particular the systems with a low-mass main-sequence companion
might have experienced angular momentum loss owing to magnetic braking
\citep{vz81}. In a recent study \citet{sg03} carefully investigated
this effect in 30 post-common-envelope binaries and found that
virtually all observed periods are close to the initial periods after
the common envelope. Only for EC 13471-1258 and BPM 71214 did they
find any evidence for significant orbital evolution. Even in these
cases the change is relatively small, so for the current purpose we
prefer to use the observed periods in the analysis.

\subsection{Double white dwarf binaries}

For double white dwarfs we can reconstruct the last phase of mass
transfer for all 19 objects listed in Table~\ref{tab:dwd}. We find
that the reconstructed values of $\alpha \lambda$ are indeed in a
reasonable range \citep[as in][]{nvy+00}. Most systems can be
explained with $\alpha \lambda \approx 0.5$.  However the spread is
large. As for the reconstructed values of $\gamma$ we again find that
all systems can be explained with a value of $\gamma \approx 1.5$.
All reconstructed values of $\gamma$ and $\alpha \lambda$ are shown in
Fig.~\ref{fig:gamma_al_wdwd_last}. As before WD1115+116 is included
twice because it is unclear which of the two objects is formed last.

%Tables of
%\begin{itemize}
%\item double white dwarfs
%\item white dwarf + M star, white dwarf + more massive
%\item sdB binaries
%\item Symbiotics, Barium stars, UV excess stars?
%\end{itemize}

\subsection{Pre-cataclysmic variables and other white dwarf -- main
  sequence stars}

\begin{figure*}
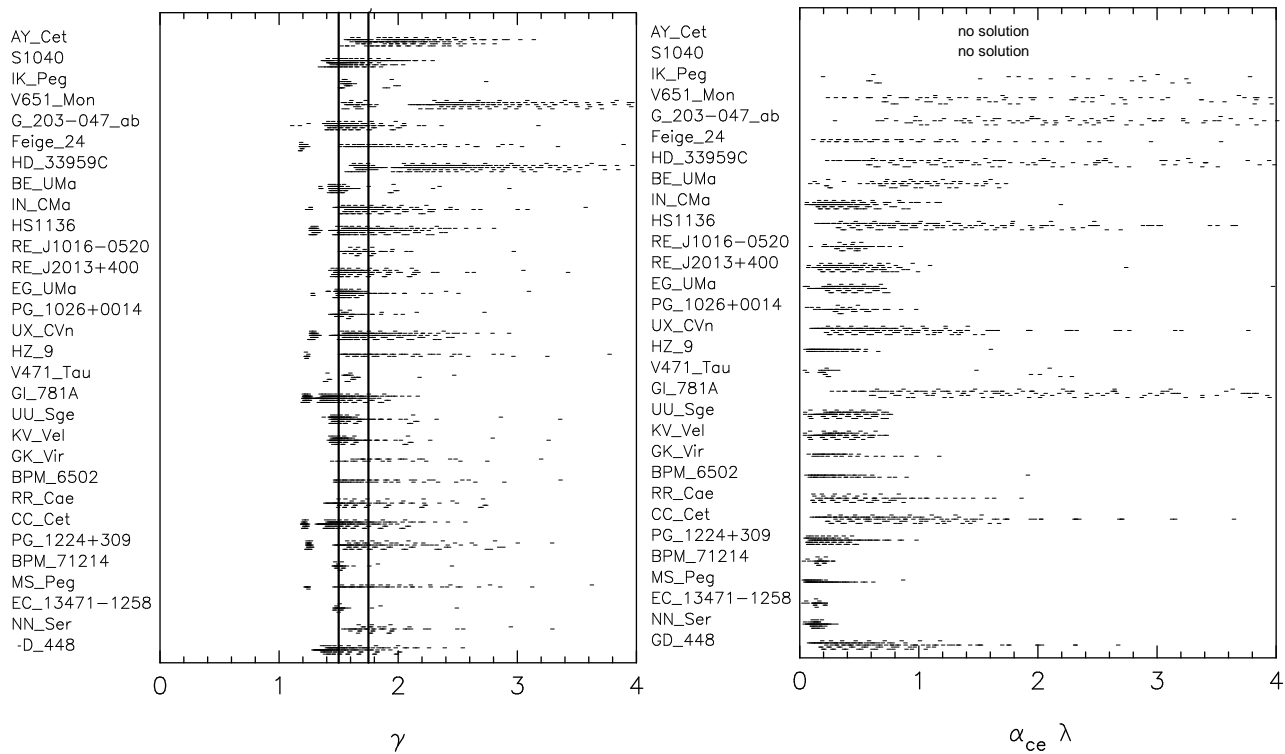

\resizebox{\columnwidth}{!}{\includegraphics[]{gamma_wddM_1211.ps}}
\resizebox{\columnwidth}{!}{\includegraphics[]{al_wddM_1211.ps}}
\caption[]{Left: Reconstructed $\gamma$ values for the mass transfer
  in white dwarf, M dwarf (or earlier type) star binaries (see
  Table~\ref{tab:wddM}). Right: reconstructed $\alpha \lambda$ values
  for the same.}
\label{fig:gamma_al_wddM}
\end{figure*}

For the properties of the observed pre-cataclysmic variables and other
white dwarf -- main-sequence binaries we use the compilation of
\citet*{hhr00} extended and updated with recent published results and
systems not in their table. All details are given in the appendix, in
Table~\ref{tab:wddM}. Most objects are short-period systems in which
the companion to the white dwarf is a low-mass main-sequence star.
Towards the bottom of the table (which is sorted by increasing orbital
period) there are a few interesting systems that have rather large
orbital periods and so would be difficult to explain with the standard
$\alpha$-formalism. Indeed in \citet{nvy+00} S1040 and AY Cet
were cited as further evidence for the \rev{$\gamma$-algorithm}.

The results of the reconstruction give a quite similar pattern to the
last phase of mass transfer in the evolution of double white dwarf
binaries: both \rev{methods} can more or less explain all the observed
systems. All values are shown in Fig.~\ref{fig:gamma_al_wddM}. As
mentioned above, at the long-period end (top of
Fig.~\ref{fig:gamma_al_wddM}), the \rev{standard $\alpha$-formalism} cannot explain a
few systems. The values of $\alpha \lambda$ also seem to correlate
with the final periods: the lower half of
Fig.~\ref{fig:gamma_al_wddM}, i.e. the shorter orbital periods,
requires lower values of $\alpha \lambda$ than the upper half.

\section{Sub-dwarf B binaries}\label{sdB_binaries}

\begin{figure*}
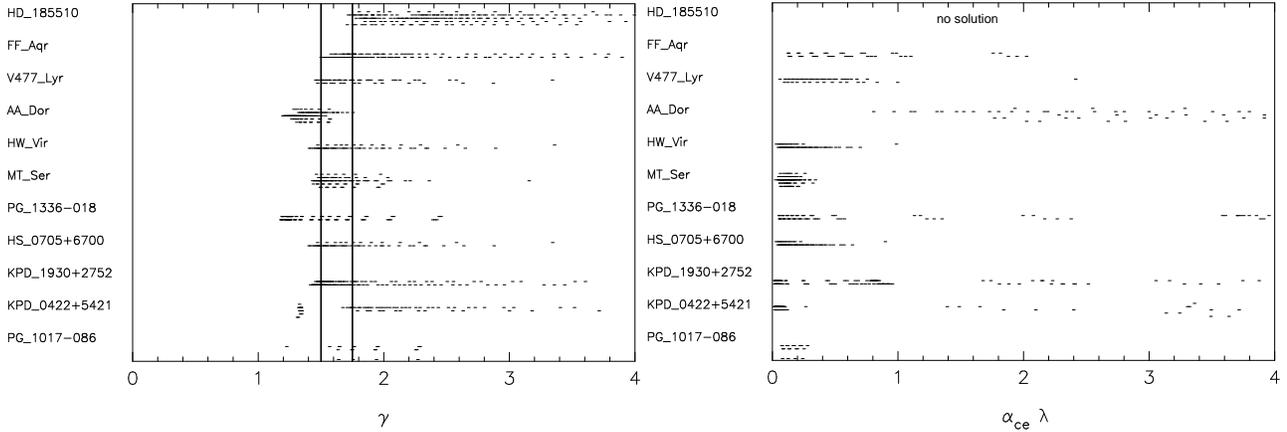

\resizebox{\columnwidth}{!}{\includegraphics[angle=-90]{gamma_sdbm2_1211.ps}}
\resizebox{\columnwidth}{!}{\includegraphics[angle=-90]{al_sdbm2_1211.ps}}
\caption[]{Left: Reconstructed $\gamma$ values for the mass transfer
  in the sdB binaries in which the mass of the companion is known (see
  Table~\ref{tab:sdBm2}).  Right: reconstructed $\alpha \lambda$
  values for the same.}
\label{fig:gamma_al_sdBm2}
\end{figure*}

\begin{figure*}
%\resizebox{2\columnwidth}{!}{\includegraphics[angle=-90]{hist.ps}}
\resizebox{\columnwidth}{!}{\includegraphics[clip]{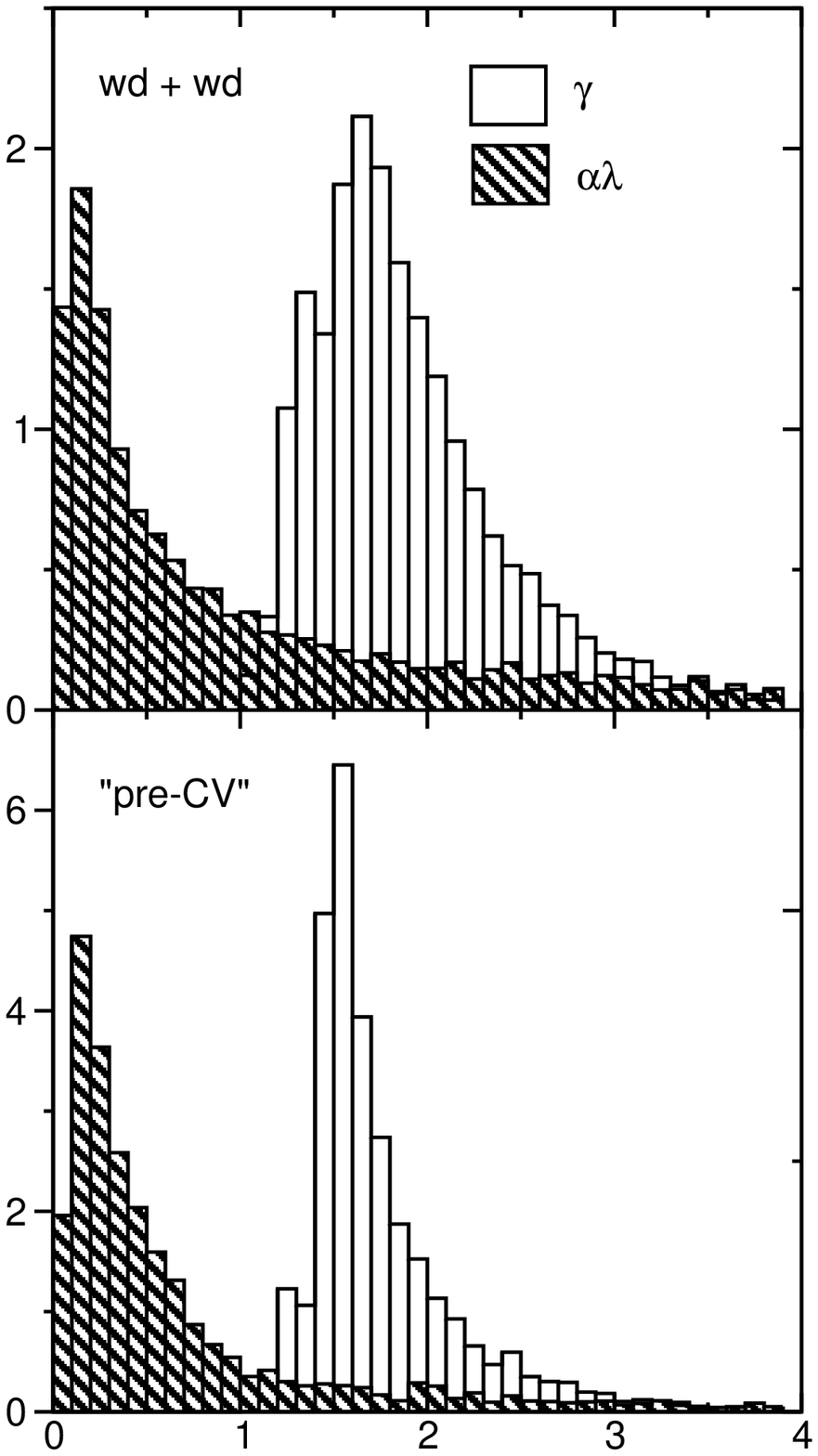}}
\resizebox{\columnwidth}{!}{\includegraphics[clip]{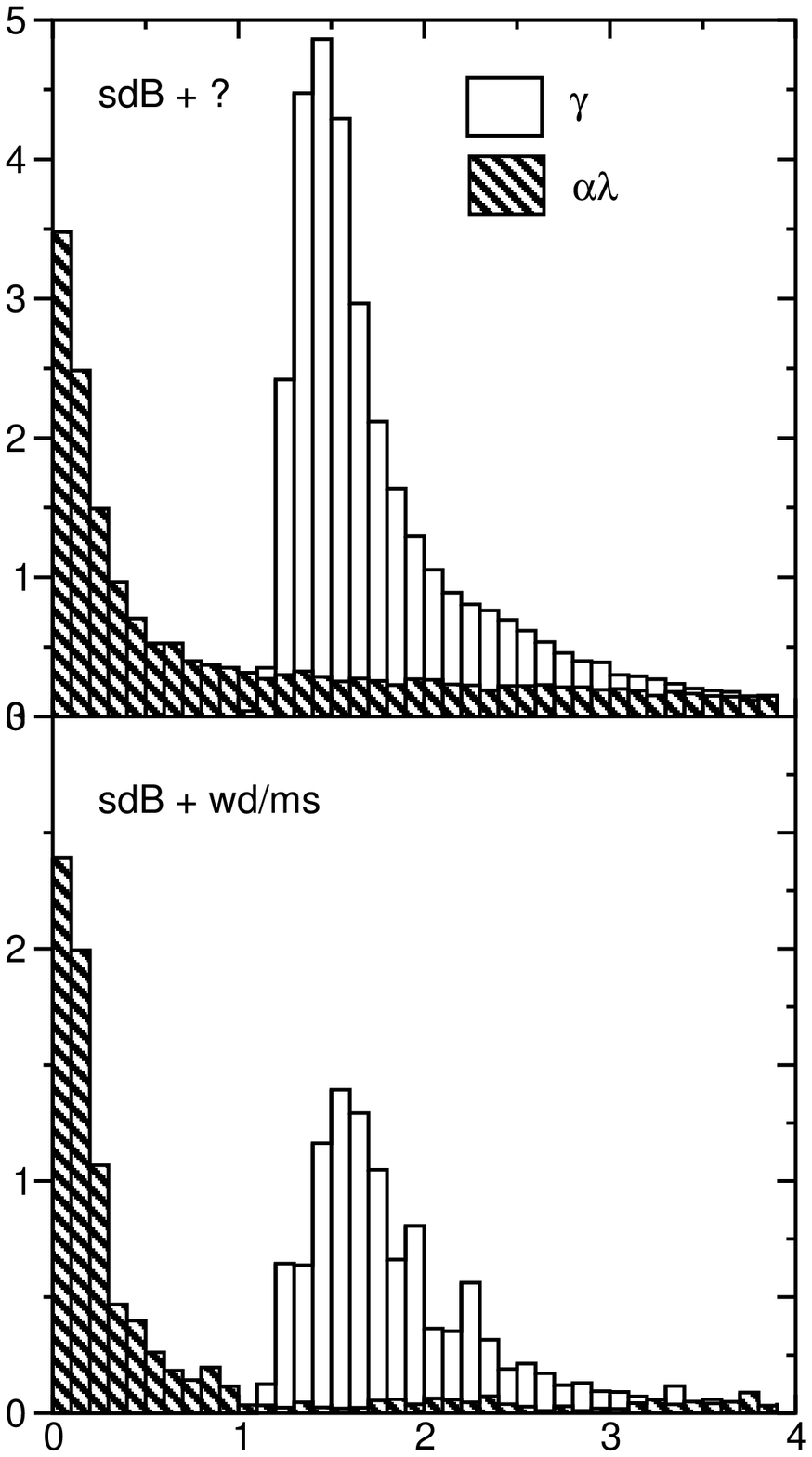}}
\caption[]{Histograms of reconstructed values of $\gamma$ and $\alpha
  \lambda$ for the last phase of mass transfer in the formation of
  double white dwarfs (top left), in white dwarf plus M dwarf binaries
  (bottom left) and the two kinds of sdB binaries (top end bottom
  right).  
}
\label{fig:hist}
\end{figure*}

The last group of binaries we consider in some detail are binaries in
which one component is a sub-dwarf B (sdB) star. These are thought to
be helium-burning stars with a very thin hydrogen envelope
\citep{heb86}.  When the core helium burning ceases, they are expected
to settle on the white dwarf cooling branch in the HR diagram. Almost
all of them are members of a binary system. There are essentially two
scenarios to form an sdB star in a binary, a giant with a
non-degenerate helium core loses its envelope to a companion
\citep[e.g.][]{hpm+02} or a a giant with a degenerate helium core
loses its envelope to a companion just before it reaches the core mass
at which the helium in the degenerate core ignites \citep[proposed
first by][ assuming the envelope was lost by a strong stellar
wind]{ddr+96}.  Detailed calculations \citep[e.g.][]{hpm+02} show that
there is a small range of core masses for which the latter occurs.
The fact that on the first giant branch the radius of the giant
increases very rapidly with the growth of the core mass, means that
this small range of core masses corresponds to a large range in radii
so no fine tuning is needed to get Roche-lobe overflow for these core
masses. The initial reasoning for this scenario came from the fact
that the observed sdB stars seemed all to be incredible similar in
mass, around 0.5 \msun. This would follow naturally from the mass at
which helium ignites (about 0.47 \msun). Note that \citet{hpm+02}
showed that the mass at helium ignition for stars initially above 1.5
\msun drops quite strongly to about 0.33 \msun at stars of initial
mass 2 \msun as the flash becomes less and less degenerate, but this
depends critically on the assumed core overshooting.

Because sdB stars are bright, they are relatively easy to study and,
in the last few years, a large fraction has been surveyed for
duplicity \citep[e.g.][]{kow98,mmm+99,mhm+01,mmm+03}. Many turn out to
be close binaries and for some the mass of the companion can also be
determined. The properties of these binaries are given in the appendix
in Table~\ref{tab:sdBm2}. We note that the general assumption is that
all sdB stars have a mass of 0.5 \msun, so only entries with a
different mass have actually been determined in detail. In
Fig.~\ref{fig:gamma_al_sdBm2} we show the reconstructed values of
$\alpha \lambda$ and $\gamma$. We find again that most systems can be
explained either by a rather low value of $\alpha \lambda$ or a value
of $\gamma$ close to 1.5 The one wide binary (HD~185510 with an
orbital period of 20.7 d) only has a solution for the
\rev{$\gamma$-algorithm}.

The sdB binaries for which the mass of the companion is not known are
listed in Table~\ref{tab:sdB}. We only list the period and the minimum
companion mass derived from the period and velocity amplitude of the
sdB stars with the assumption that the binary is seen edge on.  For
the sdB stars with no independent determination of the mass we use the
results of \citet{hpm+02} and consider all core masses that are just
about to ignite helium plus all helium core masses that are
non-degenerate as possible progenitors of the sdB stars. Since this
implies such a wide range of progenitors, almost all values of $\alpha
\lambda$ and $\gamma$ are possible. In Fig.~\ref{fig:gamma_al_sdB} we
show the reconstructed values. We use as limits to the companion mass
the minimum mass, derived from the mass function \citep{mmm+03} and
1.4 \msun because most are expected to be white dwarfs. If they are
main-sequence stars their mass should be much lower.

\section{Statistics}

Before considering the physics of the common-envelope process, we have
a look at the statistics of the reconstructed $\gamma$ and $\alpha
\lambda$ values shown in the figures in the previous sections.
Fig.~\ref{fig:hist} contains histograms of the reconstructed values of
$\gamma$ and $\alpha \lambda$ for the binaries we have considered. It
is normalised in such a way that the sum of all bins of the histograms
is equal to the total number of systems we reconstructed.  The
distribution of $\gamma$ values is actually peaked at 1.5 while the
$\alpha \lambda$ distribution is skewed to lower values. \rev{This
  suggests that the $\gamma$-algorithm is a useful tool for predicting
  the outcome of common-envelope evolution. Another} surprising result
is that each of the distributions looks very similar for the different
types of binaries.  This suggests that the processes determining the
mass transfer outcome are similar for the different types and so the
different types of binaries can be seen as independent measurements of
the same process. Furthermore it shows that the outcome of the common
envelope apparently does not depend strongly on whether the companion
star is a very compact white dwarf or a much larger main-sequence
star.

Some caution is needed in simply comparing the numerical values of
$\alpha \lambda$ and $\gamma$. The definition of $\gamma$ implies a
certain restriction to the values that can be obtained by the
reconstruction method: $\Delta J/J$ has a value between 0 and 1, and
$M_{\rm e}/(M_{\rm g} + m)$ can be written as $ M_{\rm e}/(M_{\rm g}
(1 + q))$, with $M_{\rm e}/M_{\rm g}$ limited between 0.4 and 0.9.
Combining these numbers, we simply never can find values of $\gamma$
outside the range 0 to 5. A slightly smaller range is expected for
very short period systems, where the final angular momentum is small.
That means that $\Delta J/J$ is or the order unity, leading to values
of $\gamma$ larger than 1. On the other hand, the \rev{standard
  $\alpha$-formalism} is less constrained.  Owing to its definition
($E_{\rm bind}/\Delta E_{\rm orb}$) $\alpha$ can never be zero because
the binding energy never is zero. Another property of the
\rev{standard $\alpha$-formalism} is that for most cases $\Delta
E_{\rm orb}$ is completely determined by the final separation.  The
result of that is that if the final separation is not determined by
the binding energy, giants with similar binding energy will show a
reconstructed $\alpha$ which is correlated with the final period,
which seems to be the case in Fig.~\ref{fig:gamma_al_wddM}.

\REV{We have deliberately kept the discussion focused on the
observations and tried to interpret them as model independently as
possible. The only theoretical ingredients so far are evolution models
of single stars. Though the fact that all binaries can be explained
with a single value of $\gamma$ gives a \rev{useful} tool to describe
the outcome of a common-envelope phase, it doesn't give a physical
understanding of the process. Our ideas about the interpretation of
the $\gamma$-algorithm in terms of a possible physical mechanism will
be discussed in a forthcoming paper (Nelemans \& Tout, in prep.).}

\section{Consequences of the \rev{$\gamma$-algorithm}}\label{consequences}

Although we showed above that all observed double white dwarfs,
pre-CVs and sdB binaries could be explained with the
\rev{$\gamma$-algorithm}, there are some drastic consequences of this
assumption. The most important is that the simple form of
equation~(\ref{eq:AM}) immediately shows that for given $\gamma$ there
is a limit to the amount of mass that can be lost before the system
merges. E.g. for $\gamma = 1.5$, the system will lose all its angular
momentum for $M_{\rm e} > 2/3 (M_{\rm g} + m)$. For extreme mass
ratios this is similar to a core mass fraction $\mu < 1/3$, which is
often realised in the early evolution of stars. For binaries which
undergo a first phase of stable mass transfer, the secondary often
accretes enough mass that in the second phase of mass transfer the
system will merge due to the extreme mass ratio. Another situation
where a large fraction of the total mass is lost is in the case of
mass transfer from a giant to another giant leading to a double
spiral-in \citep[e.g.][]{bro95,nyp+00}. If in that situation the
\rev{$\gamma$-algorithm} is used, it normally leads to complete
merger of the two cores.

\subsection{Double white dwarfs and (pre-)CVs.}\label{popsynth}

In order to assess the consequences of using the
\rev{$\gamma$-algorithm} in all phases of dynamically unstable mass
transfer, we made a population synthesis calculation with all
assumptions identical to the model described in \citet*{nyp03}, except
for the common-envelope phase, where we now use the
\rev{$\gamma$-algorithm} in all cases.

For double white dwarfs the main effect is that the number of
observable systems in the Galaxy goes down by almost a factor of 3.
On the one hand this is due to the fact that more systems merge
because all the angular momentum is lost. This happens for systems in
which the secondary is rather massive by itself or because it accretes
in the first phase of mass transfer, or because mass transfer begins
when both stars are giants. On the other hand more systems form with
such long orbital periods (above 40 d) so that current observing
programs are not sensitive to them. This happens for the the systems
forming from low-mass stars.

The period distribution of the model in which we use the standard
$\alpha$-formalism in the last phase of mass transfer matches the observed
distribution quite well \citep{nyp+00} so we would expect the new
model to do less well. However, this is not significantly the case.
The reason is that, because of the relatively small number of observed
systems, we are essentially comparing the observed period range with
that in the model. The short end of this range is largely determined
by the fact that systems merge due to angular momentum loss by
gravitational wave radiation and thus disappear from the observable
sample and the long period end by the limits of the current methods of
period determinations.

The difference in the model is very important for the merger rate of
double white dwarfs. It drops by about a factor 6.  For massive pairs
the situation is even more dramatic. The merger rate of pairs with a
mass above the Chandrasekhar mass reduces from $1.1 \times 10^{-3}
{\rm yr}^{-1}$ to $7.7 \times 10^{-6} {\rm yr}^{-1}$.  Similarly, the
confusion limited noise background of the Galactic double white dwarf
population for the space based gravitational wave detector
\textit{LISA} decreases by about a factor 2.

An even greater change is that the mass-ratio distribution of the
close double white dwarfs is even more peaked around unity when the
$\gamma$-model is used throughout. This means the chances for
double white dwarfs to start stable mass transfer and evolve into AM
CVn systems are significantly reduced. The birth rate of AM CVn
systems drops by a factor 20 from $1.3 \times 10^{-3} {\rm yr}^{-1}$
to $6.4 \times 10^{-5} {\rm yr}^{-1}$, while the total number of
systems in the Galaxy goes from $2.3 \times 10^7$ to $9.2 \times
10^5$. For AM CVn systems formed from helium stars \citep[see][for a
discussion of the ways to form AM CVn systems]{npv+00} the reduction
is even larger, about a factor of 100.  With such a small number it
would become problematic to explain the number of known systems which
are believed to be only a small part of the total observable
population \citep[e.g.][]{npv+00}.

As expected from the fact that we can explain most observed white
dwarf -- main-sequence binaries, this population is not much affected
by the use of the \rev{$\gamma$-algorithm}. \REV{Of course the white
dwarfs with relatively massive companions have longer periods compared
to the case when we use the $\alpha$-formalism in the first phase of
mass transfer \citep[see][]{nvy+00}.} There is quite a large effect on
the formation of cataclysmic variables, with their current birth rate
reduced by a factor 2.5, but the total number in the Galaxy by a
factor 7 (from 26.8 to 3.8 million). However, these numbers are within
the range expected from observations \citep[e.g.][]{mar01b},
particularly because, in our Galactic model, the fraction of systems
close to the sun is lower than in an exponential disc
\citep[see][]{nyp03}.

\subsection{Consequences for other binaries}

Some of the properties of the \rev{$\gamma$-algorithm} are relevant
to the formation of symbiotic stars, barium and S-stars
\citep[e.g.][]{jvm+98}. As shown in Fig.~\ref{fig:Mc_m}, it can lead
to quite wide binaries after unstable mass transfer. otherwise the
observed long periods of barium stars mean that population models rely
on either wind accretion to transfer s-process enriched material
\citep*[e.g.][]{bj88,ktl00} or extra mass loss on the AGB to avoid a
common envelope \citep{te88,hep+95}. A detailed analysis of the
post-AGB binary in the Red Rectangle \citep{mst+02} suggests an
evolutionary scenario in which the $\gamma$-mechanism is
needed to explain the current system parameters.

Similar problems also affect higher-mass binaries, particularly the
common-envelope phases in the evolution leading to low-mass X-ray
binaries and double neutron stars.  The standard scenario for the
formation of low-mass X-ray binaries \citep{heu83} involves the
common-envelope evolution of a star that, after losing its envelope,
becomes a neutron star or even black hole and a low-mass main-sequence
star. A relatively low-mass neutron star progenitor with an initial
mass of 9 \msun, which attains a maximum core mass of about 2.5
\msun when its total mass is 8.5 \msun, according to the \citet{hpt00}
formulae, has $M_{\rm e}/ (M_{\rm g} + m) > 0.63$ and so must have
$\gamma < 1.58$ in order not to merge. It turns out that for more
massive stars this limit on $\gamma$ decreases only rather slowly, so
for values of $\gamma$ not too much in excess of 1.5 formation of
low-mass X-ray binaries is still possible.  For the formation of
double neutron stars the situation is quite similar.

\section{Conclusions}\label{conclusions}

We have used the masses of observed white dwarfs in binaries to
estimate the radii of their progenitors on the assumption that the
white dwarf masses are good approximations to the core masses of their
progenitor giants.  Using these progenitor masses we have
reconstructed the parameters of the progenitor binary systems of the
observed white dwarf binaries.  These, together with the observed
binary parameters, were used to reconstruct the change in orbital
separation during the mass transfer phase in which the white dwarf was
formed. By comparing this change to the expected change for the
\rev{standard $\alpha$-formalism}, \rev{explicitly} based on the
energy balance, and an the \rev{$\gamma$-algorithm}, \rev{explicitly}
based on the angular momentum balance, we derived the values of the
free parameters in these \rev{methods}.

The main result is that, as was found earlier, for the first phase of
mass transfer in the evolution leading to the currently observed
double white dwarf systems, the \rev{standard $\alpha$-formalism}
cannot explain the observations, while the \rev{$\gamma$-algorithm}
can. For all the other reconstructed phases either \rev{method} can
explain the observations.  However, the reconstructed values for the
\rev{$\gamma$-algorithm} strongly cluster around 1.5, while the values
of the free parameter in the \rev{standard $\alpha$-formalism} (the
efficiency parameter), show a wide range of values, skewed towards low
($\la 0.5$) values. \rev{Thus the predictive power, at least in
  statistical sense, of the $\gamma$-algorithm seems to be greater
  than the standard $\alpha$-formalism}.

%As a first step towards a physical interpretation of our results, we
%considered a hybrid picture, in which the initial phase of the
%common-envelope evolution is driven by the angular momentum in the
%system, and only the last phase for systems that do not have enough
%orbital angular momentum to spin-up the giant star in the binary, is
%governed by the \rev{standard $\alpha$-formalism}. \rev{Although the
%  plausibility of such a hybrid picture is very uncertain, it} seems
%to be able to explain all observed binaries.

\section*{Acknowledgments}

We thank Philipp Podsiadlowski for useful discussions \rev{and the
  referee for comments that improved the presentation of the paper.}
GN is supported by PPARC.  CAT thanks Churchill College for a
fellowship. This research has made extensive use of NASA's
Astrophysics Data System.

\bibliography{journals,binaries}

\begin{thebibliography}{74}
\expandafter\ifx\csname natexlab\endcsname\relax\def\natexlab#1{#1}\fi

\bibitem[{{Benedict} et~al.(2000)}]{bmf+00}
{Benedict}, G.~F., et~al., 2000, \aj, 119, 2382

\bibitem[{{Bergeron} et~al.(1989){Bergeron}, {Wesemael}, {Fontaine}, \&
  {Liebert}}]{bwf+89}
{Bergeron}, P., {Wesemael}, F., {Fontaine}, G., {Liebert}, J., 1989, \apjl,
  345, L91

\bibitem[{{Bleach} et~al.(2000){Bleach}, {Wood}, {Catal{\' a}n}, {Welsh},
  {Robinson}, \& {Skidmore}}]{bwc+00}
{Bleach}, J.~N., {Wood}, J.~H., {Catal{\' a}n}, M.~S., {Welsh}, W.~F.,
  {Robinson}, E.~L., {Skidmore}, W., 2000, \mnras, 312, 70

\bibitem[{{Boffin} \& {Jorissen}(1988)}]{bj88}
{Boffin}, H.~M.~J., {Jorissen}, A., 1988, \aap, 205, 155

\bibitem[{Bragaglia et~al.(1990)Bragaglia, Greggio, Renzini, \&
  D'Odorico}]{bgr+90}
Bragaglia, A., Greggio, L., Renzini, A., D'Odorico, S., 1990, ApJ, 365, L13

\bibitem[{Brown(1995)}]{bro95}
Brown, G.~E., 1995, \apj, 440, 270

\bibitem[{D'Cruz et~al.(1996)D'Cruz, Dorman, Rood, \& O'Connell}]{ddr+96}
D'Cruz, N.~L., Dorman, B., Rood, R.~T., O'Connell, R.~W., 1996, ApJ, 466, 359

\bibitem[{de~Kool \& Ritter(1993)}]{kr93}
de~Kool, M., Ritter, H., 1993, \aap, 267, 397

\bibitem[{de~Kool et~al.(1987)de~Kool, van~den Heuvel, \& Pylyser}]{khp87}
de~Kool, M., van~den Heuvel, E. P.~J., Pylyser, E., 1987, A\&A, 183, 47

\bibitem[{{Delfosse} et~al.(1999){Delfosse}, {Forveille}, {Beuzit}, {Udry},
  {Mayor}, \& {Perrier}}]{dfb+99}
{Delfosse}, X., {Forveille}, T., {Beuzit}, J.-L., {Udry}, S., {Mayor}, M.,
  {Perrier}, C., 1999, \aap, 344, 897

\bibitem[{{Drechsel} et~al.(2001)}]{dhn+01}
{Drechsel}, H., et~al., 2001, \aap, 379, 893

\bibitem[{Eggleton(1983)}]{egg83}
Eggleton, P.~P., 1983, ApJ, 268, 368

\bibitem[{{Gizis}(1998)}]{giz98}
{Gizis}, J.~E., 1998, \aj, 115, 2053

\bibitem[{Han(1998)}]{han98}
Han, Z., 1998, MNRAS, 296, 1019

\bibitem[{{Han} et~al.(1995){Han}, {Eggleton}, {Podsiadlowski}, \&
  {Tout}}]{hep+95}
{Han}, Z., {Eggleton}, P.~P., {Podsiadlowski}, P., {Tout}, C.~A., 1995, \mnras,
  277, 1443

\bibitem[{{Han} et~al.(2002){Han}, {Podsiadlowski}, {Maxted}, {Marsh}, \&
  {Ivanova}}]{hpm+02}
{Han}, Z., {Podsiadlowski}, P., {Maxted}, P.~F.~L., {Marsh}, T.~R., {Ivanova},
  N., 2002, \mnras, 336, 449

\bibitem[{{Heber}(1986)}]{heb86}
{Heber}, U., 1986, \aap, 155, 33

\bibitem[{{Hilditch} et~al.(1996){Hilditch}, {Harries}, \& {Hill}}]{hhh96}
{Hilditch}, R.~W., {Harries}, T.~J., {Hill}, G., 1996, \mnras, 279, 1380

\bibitem[{Hillwig et~al.(2000)Hillwig, Honeycutt, \& Robertson}]{hhr00}
Hillwig, T.~C., Honeycutt, R.~K., Robertson, J.~W., 2000, \aj, 120, 1113

\bibitem[{{Holberg} et~al.(1995){Holberg}, {Saffer}, {Tweedy}, \&
  {Barstow}}]{hst+95}
{Holberg}, J.~B., {Saffer}, R.~A., {Tweedy}, R.~W., {Barstow}, M.~A., 1995,
  \apjl, 452, L133

\bibitem[{Hurley et~al.(2000)Hurley, Pols, \& Tout}]{hpt00}
Hurley, J.~R., Pols, O.~R., Tout, C.~A., 2000, \mnras, 315, 543

\bibitem[{{Hut}(1980)}]{hut80}
{Hut}, P., 1980, \aap, 92, 167

\bibitem[{Iben et~al.(1997)Iben, Tutukov, \& Yungelson}]{ity97}
Iben, Jr, I., Tutukov, A.~V., Yungelson, L.~R., 1997, ApJ, 475, 291

\bibitem[{{Jeffery} \& {Simon}(1997)}]{js97}
{Jeffery}, C.~S., {Simon}, T., 1997, \mnras, 286, 487

\bibitem[{{Jorissen} et~al.(1998){Jorissen}, {Van Eck}, {Mayor}, \&
  {Udry}}]{jvm+98}
{Jorissen}, A., {Van Eck}, S., {Mayor}, M., {Udry}, S., 1998, \aap, 332, 877

\bibitem[{{Karakas} et~al.(2000){Karakas}, {Tout}, \& {Lattanzio}}]{ktl00}
{Karakas}, A.~I., {Tout}, C.~A., {Lattanzio}, J.~C., 2000, \mnras, 316, 689

\bibitem[{{Karl} et~al.(2002){Karl}, {Napiwotzki}, {Heber}, {Lisker},
  {Nelemans}, \& {Reimers}}]{knh+02}
{Karl}, C., {Napiwotzki}, R., {Heber}, U., {Lisker}, T., {Nelemans},
  G.~{Christlieb}, N., {Reimers}, 2002, in de~Martino, D., Kalytis, R.,
  Silvotti, R., Solheim, J., eds., White Dwarfs, Proc. XIII Workshop on White
  Dwarfs, Kluwer, p.~43

\bibitem[{Karl et~al.(2003)Karl, {Napiwotzki}, {Nelemans}, {Christlieb},
  {Koester}, {Heber}, \& {Reimers}}]{knn03}
Karl, C., {Napiwotzki}, R., {Nelemans}, G., {Christlieb}, N., {Koester}, D.,
  {Heber}, U., {Reimers}, D., 2003, \aap, 410, 663

\bibitem[{{Kawka} et~al.(2000){Kawka}, {Vennes}, {Dupuis}, \& {Koch}}]{kvd+00}
{Kawka}, A., {Vennes}, S., {Dupuis}, J., {Koch}, R., 2000, \aj, 120, 3250

\bibitem[{{Kawka} et~al.(2002){Kawka}, {Vennes}, {Koch}, \&
  {Williams}}]{kvk+02}
{Kawka}, A., {Vennes}, S., {Koch}, R., {Williams}, A., 2002, \aj, 124, 2853

\bibitem[{{Kilkenny} et~al.(1998){Kilkenny}, {O'Donoghue}, {Koen},
  {Lynas-Gray}, \& {van Wyk}}]{kok+98}
{Kilkenny}, D., {O'Donoghue}, D., {Koen}, C., {Lynas-Gray}, A.~E., {van Wyk},
  F., 1998, \mnras, 296, 329

\bibitem[{Koen et~al.(1998)Koen, Orosz, \& Wade}]{kow98}
Koen, C., Orosz, J.~A., Wade, R.~A., 1998, MNRAS, 300, 695

\bibitem[{Landsman et~al.(1997)Landsman, Aparicio, Bergeron, Di~Stefano, \&
  Stecher}]{lab+97}
Landsman, W., Aparicio, J., Bergeron, P., Di~Stefano, R., Stecher, T.~P., 1997,
  ApJ, 481, L93

\bibitem[{Marsh(1995)}]{mar95}
Marsh, T.~R., 1995, MNRAS, 275, L1

\bibitem[{Marsh(2000)}]{mar00}
Marsh, T.~R., 2000, New Astronomy Review, 44, 119

\bibitem[{{Marsh}(2001)}]{mar01b}
{Marsh}, T.~R., 2001, in Vanbeveren, D., ed., The Influence of Binaries on
  Stellar Population Studies, vol. 264 of \emph{ASSL}, Kluwer, Dordrecht, p.~55

\bibitem[{Marsh et~al.(1995)Marsh, Dhillon, \& Duck}]{mdd95}
Marsh, T.~R., Dhillon, V.~S., Duck, S.~R., 1995, MNRAS, 275, 828

\bibitem[{Maxted \& Marsh(1999)}]{mm99}
Maxted, P. F.~L., Marsh, T.~R., 1999, MNRAS, 307, 122

\bibitem[{Maxted et~al.(2000{\natexlab{a}})Maxted, Marsh, Moran, \&
  Han}]{mmm+00}
Maxted, P. F.~L., Marsh, T.~R., Moran, C. K.~J., Han, Z., 2000{\natexlab{a}},
  MNRAS, 314, 334

\bibitem[{Maxted et~al.(2000{\natexlab{b}})Maxted, Marsh, \& North}]{mmn00}
Maxted, P. F.~L., Marsh, T.~R., North, R.~C., 2000{\natexlab{b}}, MNRAS, 317,
  L41

\bibitem[{Maxted et~al.(2001)Maxted, Heber, Marsh, \& North}]{mhm+01}
Maxted, P.~F.~L., Heber, U., Marsh, T.~R., North, R.~C., 2001, \mnras, 326,
  1391

\bibitem[{{Maxted} et~al.(2002{\natexlab{a}}){Maxted}, {Burleigh}, {Marsh}, \&
  {Bannister}}]{mbm+02}
{Maxted}, P.~F.~L., {Burleigh}, M.~R., {Marsh}, T.~R., {Bannister}, N.~P.,
  2002{\natexlab{a}}, \mnras, 334, 833

\bibitem[{{Maxted} et~al.(2002{\natexlab{b}}){Maxted}, {Marsh}, {Heber},
  {Morales-Rueda}, {North}, \& {Lawson}}]{mmh+02}
{Maxted}, P.~F.~L., {Marsh}, T.~R., {Heber}, U., {Morales-Rueda}, L., {North},
  R.~C., {Lawson}, W.~A., 2002{\natexlab{b}}, \mnras, 333, 231

\bibitem[{{Maxted} et~al.(2002{\natexlab{c}}){Maxted}, {Marsh}, \&
  {Moran}}]{mmm02}
{Maxted}, P.~F.~L., {Marsh}, T.~R., {Moran}, C.~K.~J., 2002{\natexlab{c}},
  \mnras, 332, 745

\bibitem[{{Men'shchikov} et~al.(2002){Men'shchikov}, {Schertl}, {Tuthill},
  {Weigelt}, \& {Yungelson}}]{mst+02}
{Men'shchikov}, A.~B., {Schertl}, D., {Tuthill}, P.~G., {Weigelt}, G.,
  {Yungelson}, L.~R., 2002, \aap, 393, 867

\bibitem[{{Morales-Rueda} et~al.(2003){Morales-Rueda}, {Maxted}, {Marsh},
  {North}, \& {Heber}}]{mmm+03}
{Morales-Rueda}, L., {Maxted}, P.~F.~L., {Marsh}, T.~R., {North}, R.~C.,
  {Heber}, U., 2003, \mnras, 338, 752

\bibitem[{Moran et~al.(1999)Moran, Maxted, Marsh, Saffer, \& Livio}]{mmm+99}
Moran, C., Maxted, P., Marsh, T.~R., Saffer, R.~A., Livio, M., 1999, MNRAS,
  304, 535

\bibitem[{Moran et~al.(1997)Moran, Marsh, \& Bragaglia}]{mmb97}
Moran, C. K.~J., Marsh, T.~R., Bragaglia, A., 1997, MNRAS, 288, 538

\bibitem[{{Napiwotzki} et~al.(2001)}]{ncd+01}
{Napiwotzki}, R., et~al., 2001, Astronomische Nachrichten, 322, 411

\bibitem[{{Napiwotzki} et~al.(2002{\natexlab{a}})}]{nkn02}
{Napiwotzki}, R., et~al., 2002{\natexlab{a}}, \aap, 386, 957

\bibitem[{{Napiwotzki} et~al.(2002{\natexlab{b}})}]{ndh+02}
{Napiwotzki}, R., et~al., 2002{\natexlab{b}}, in de~Martino, D., Kalytis, R.,
  Silvotti, R., Solheim, J., eds., White Dwarfs, Proc. XIII Workshop on White
  Dwarfs, Kluwer, p.~39

\bibitem[{Nelemans \& Tauris(1998)}]{nt98}
Nelemans, G., Tauris, T.~M., 1998, A\&A, 335, L85

\bibitem[{Nelemans et~al.(2000)Nelemans, Verbunt, Yungelson, \&
  Portegies~Zwart}]{nvy+00}
Nelemans, G., Verbunt, F., Yungelson, L.~R., Portegies~Zwart, S.~F., 2000,
  A\&A, 360, 1011

\bibitem[{Nelemans et~al.(2001{\natexlab{a}})Nelemans, Portegies~Zwart,
  Verbunt, \& Yungelson}]{npv+00}
Nelemans, G., Portegies~Zwart, S.~F., Verbunt, F., Yungelson, L.~R.,
  2001{\natexlab{a}}, A\&A, 368, 939

\bibitem[{Nelemans et~al.(2001{\natexlab{b}})Nelemans, Yungelson,
  Portegies~Zwart, \& Verbunt}]{nyp+00}
Nelemans, G., Yungelson, L.~R., Portegies~Zwart, S.~F., Verbunt, F.,
  2001{\natexlab{b}}, A\&A, 365, 491

\bibitem[{Nelemans et~al.(2004)Nelemans, Yungelson, \& Portegies~Zwart}]{nyp03}
Nelemans, G., Yungelson, L.~R., Portegies~Zwart, S.~F., 2004, \mnras, 349, 181

\bibitem[{{O'Brien} et~al.(2001){O'Brien}, {Bond}, \& {Sion}}]{obs01}
{O'Brien}, M.~S., {Bond}, H.~E., {Sion}, E.~M., 2001, \apj, 563, 971

\bibitem[{{O'Donoghue} et~al.(2003){O'Donoghue}, {Koen}, {Kilkenny}, {Stobie},
  {Koester}, {Bessell}, {Hambly}, \& {MacGillivray}}]{okk+03}
{O'Donoghue}, D., {Koen}, C., {Kilkenny}, D., {Stobie}, R.~S., {Koester}, D.,
  {Bessell}, M.~S., {Hambly}, N., {MacGillivray}, H., 2003, \mnras, 345, 506

\bibitem[{Orosz \& Wade(1999)}]{ow00}
Orosz, J.~A., Wade, R.~A., 1999, MNRAS, 310, 773

\bibitem[{Paczy\'nski(1976)}]{pac76}
Paczy\'nski, B., 1976, in Eggleton, P., Mitton, S., Whelan, J., eds., Structure
  and Evolution of Close Binary Systems, Kluwer, Dordrecht, p.~75

\bibitem[{{Rauch}(2000)}]{rau00}
{Rauch}, T., 2000, \aap, 356, 665

\bibitem[{Saffer et~al.(1988)Saffer, Liebert, \& Olszewski}]{slo88}
Saffer, R.~A., Liebert, J., Olszewski, E.~W., 1988, ApJ, 334, 947

\bibitem[{{Saffer} et~al.(1993){Saffer}, {Wade}, {Liebert}, {Green}, {Sion},
  {Bechtold}, {Foss}, \& {Kidder}}]{swl+93}
{Saffer}, R.~A., {Wade}, R.~A., {Liebert}, J., {Green}, R.~F., {Sion}, E.~M.,
  {Bechtold}, J., {Foss}, D., {Kidder}, K., 1993, \aj, 105, 1945

\bibitem[{{Schreiber} \& {G{\" a}nsicke}(2003)}]{sg03}
{Schreiber}, M.~R., {G{\" a}nsicke}, B.~T., 2003, \aap, 406, 305

\bibitem[{Simon et~al.(1985)Simon, Fekel, \& Gibson~Jr}]{sfg85}
Simon, T., Fekel, F.~C., Gibson~Jr, D.~M., 1985, ApJ, 295, 153

\bibitem[{{Sparks} \& {Stecher}(1974)}]{ss74}
{Sparks}, W.~M., {Stecher}, T.~P., 1974, \apj, 188, 149

\bibitem[{Tout \& Eggleton(1988)}]{te88}
Tout, C.~A., Eggleton, P.~P., 1988, MNRAS, 231, 823

\bibitem[{van~den Heuvel(1983)}]{heu83}
van~den Heuvel, E. P.~J., 1983, in Lewin, W. H.~G., van~den Heuvel E. P.~J.,
  eds., Accretion-driven stellar X-ray sources, CUP, Cambridge, p. 303

\bibitem[{{Vennes} et~al.(1998){Vennes}, {Christian}, \& {Thorstensen}}]{vct98}
{Vennes}, S., {Christian}, D.~J., {Thorstensen}, J.~R., 1998, \apj, 502, 763

\bibitem[{{Vennes} et~al.(1999){Vennes}, {Thorstensen}, \& {Polomski}}]{vtp99}
{Vennes}, S., {Thorstensen}, J.~R., {Polomski}, E.~F., 1999, \apj, 523, 386

\bibitem[{Verbunt \& Zwaan(1981)}]{vz81}
Verbunt, F., Zwaan, C., 1981, A\&A, 100, L7

\bibitem[{Webbink(1984)}]{web84}
Webbink, R.~F., 1984, ApJ, 277, 355

\bibitem[{{Wood} \& {Saffer}(1999)}]{ws99}
{Wood}, J.~H., {Saffer}, R., 1999, \mnras, 305, 820

\bibitem[{Zahn(1977)}]{zah77}
Zahn, J.-P., 1977, A\&A, 57, 383, erratum 1978, A\&A 67, 162

\end{thebibliography}
\bibliographystyle{mn_new}

\label{lastpage}

\appendix

\section{Data and graphs white dwarf and \lowercase{sd}B binaries}\label{appendix}

\begin{table}
\caption[]{Properties of the observed white dwarf, other star binaries}
\label{tab:wddM}
\begin{tabular}{ldddr} \hline
Object    & \multicolumn{1}{c}{$P$} & \multicolumn{1}{c}{$M_{\rm WD}$} & \multicolumn{1}{c}{$m$} &  Ref \\ 
  & \multicolumn{1}{c}{(d)} & \multicolumn{1}{c}{(\msun)} &
    \multicolumn{1}{c}{(\msun)} & \\\hline
GD 448         &0.103042   &0.41    &0.096 & 1\\
NN Ser         &0.130080   &0.570   &0.120 & 1\\
EC 13471$-$1258&0.15074    &0.78    &0.43  & 10 \\
MS Peg         &0.173660   &0.480   &0.220 & 1\\
BPM 71214      &0.20162    &0.77    &0.4   & 9\\
PG 1224+309    &0.25869    &0.45    &0.28  & 1 \\
CC Cet         &0.286654   &0.400   &0.180 & 1\\
RR Cae         &0.304      &0.467   &0.095 & 1 \\
BPM 6502       &0.33678    &0.5     & 0.15 & 8 \\
GK Vir         &0.344331   &0.510   &0.100 & 1\\
KV Vel         &0.357113   &0.630   &0.25  & 11 \\
UU Sge         &0.465069   &0.630   &0.290 & 1\\
Gl 781A        &0.497      &0.35    &0.25  & 4        \\
V471 Tau       &0.521183   &0.84    &0.93  &   2\\
HZ 9           &0.564330   &0.510   &0.280 & 1\\
UX CVn         &0.573703   &0.390   &0.420 & 1\\
PG 1026 0014   &0.597257   &0.65    &0.220 & 15\\
EG UMa         &0.667650   &0.64    &0.42  & 13\\
RE J2013$+$400 &0.7056     &0.56    &0.18  & 12\\
RE J1016$-$0520&0.789      &0.61    &0.15  & 12 \\
IN CMa         &1.262450   &0.57    &0.39  & 12\\
BE UMa         &2.291166   &0.7     &0.36  &  1\\
HD 33959C      &2.99       &0.6     &1.5   & 14\\
Feige 24       &4.231600   &0.49    &0.37  & 13 \\
G 203$-$047ab  &14.7136    &0.6     &0.33  & 3  \\
V651 Mon       &15.991000  &0.400   &1.800 & 1\\
IK Peg         &21.721700  &1.100   &1.700 & 1\\
S1040          &42.8       &0.22    &1.5   &  6\\
AY Cet         &56.8       &0.25    &2.2   &  7\\ \hline
\end{tabular}

References:\\
(1) \citet*{hhr00};
(2) \citet*{obs01};
(3) \citet{dfb+99};
(4) \citet{giz98};
(5) \citet{bmf+00};
(6) \citet{lab+97};
(7) \citet*{sfg85};
(8) \citet{kvd+00};
(9) \citet{kvk+02};
(10) \citet{okk+03};
(11) \citet*{hhh96};
(12) \citet*{vtp99};
(13) \citet{bwc+00}
(14) \citet*{vct98}, note that \citet{hhr00} list the parameters of HD~33959A;
(15) \citet{swl+93}
\end{table}

\begin{table}
\caption[]{Properties of the observed sdB binaries, for which the mass
of the companion is known.}
\label{tab:sdBm2}
\begin{tabular}{ldddr} \hline
Object    & \multicolumn{1}{c}{$P$} & \multicolumn{1}{c}{$m$} & \multicolumn{1}{c}{$M$} &  Ref \\ 
  & \multicolumn{1}{c}{(d)} & \multicolumn{1}{c}{(\msun)} &
    \multicolumn{1}{c}{(\msun)} & \\\hline
PG 1017$-$086     &0.073      &0.5     &0.078  &9 \\
KPD 0422$+$5421   &0.090180   &0.51    &0.526  & 4 \\
KPD 1930$+$2752   &0.095111   &0.5     &0.97   & 7\\
HS0 705$+$6700    &0.095647   &0.48    &0.134  & 8\\
PG 1336$-$018     &0.101      &0.5     &0.15   & 3\\
MT Ser         &0.113227   &0.6   &0.2     & 1\\
HW Vir         &0.116720   &0.48   &0.14   & 5\\
AA Dor         &0.261540   &0.330   &0.066 &  6\\
V477 Lyr       &0.471729   &0.51   &0.15   & 1\\
FF Aqr         &9.207755   &0.5    &2.0    & 1\\
HD 185510     &20.7        &0.304  &2.27   & 2\\ \hline
\end{tabular}
References:\\
(1) \citet{hhr00};
(2) \citet{js97};
(3) \citet{kok+98};
(4) \citet{ow00};
(5) \citet{ws99};
(6) \citet{rau00};
(7) \citet*{mmn00};
(8) \citet{dhn+01};
(9) \citet{mmh+02}
\end{table}

\begin{table}
\caption[]{Periods and minimum companion masses of sdB binaries, from \citet{mmm+03}}
\label{tab:sdB}
\begin{tabular}{ldd} \hline
Object    &   \multicolumn{1}{c}{$P$} & \multicolumn{1}{c}{$M_{\rm min}$} \\ 
  & \multicolumn{1}{c}{(d)} & \multicolumn{1}{c}{(\msun)} \\\hline
PG1017$-$086 &   0.0729939 &   0.066 \\
PG1043$+$760 &   0.1201506 &   0.106 \\
PG1432$+$159 &   0.22489 &     0.294 \\
PG2345$+$318 &   0.2409458 &   0.379 \\
PG1329$+$159 &   0.249699 &    0.083 \\
TW Crv &       0.328 &       - \\
PG1101$+$249 &   0.35386 &     0.424 \\
KPD1946$+$4340 & 0.403739 &    0.628 \\
PG1743$+$477 &   0.515561 &    0.438 \\
PG0001$+$275 &   0.528 &       0.293 \\
PG0101$+$039 &   0.569908 &    0.370  \\
PG1725$+$252 &   0.601507 &    0.381 \\
PG1247$+$554 &   0.602740 &    0.090 \\
PG1248$+$164 &   0.73232 &     0.207 \\
PG0849$+$319 &   0.7451 &      0.228 \\
PG1627$+$017 &   0.829226 &    0.273 \\
PG1116$+$301 &   0.85621 &     0.356 \\
PG0918$+$029 &   0.87679 &     0.313 \\
HE1047$-$0436 &  1.213253 &    0.458 \\
PG0133$+$114 &   1.2382 &      0.388 \\
PG1512$+$244 &   1.26978 &     0.458 \\
UVO1735$+$22 &   1.278 &      0.539 \\
HD 171858    &   1.529   &    0.510 \\
PG1716$+$426 &   1.77732 &     0.366 \\
PG1300$+$279 &   2.2593 &      0.346 \\
PG1538$+$269 &   2.501 &       0.600 \\
KPD0025$+$5402 & 3.571 &       0.235 \\
PG0839$+$399 &   5.622 &       0.226 \\
PG0907$+$123 &   6.1163 &      0.521 \\
PG1032$+$406 &   6.779 &       0.247 \\
PG0940$+$068 &   8.330 &       0.634 \\
PG1110$+$294 &   9.415 &       0.633 \\
PG1619$+$522 &  15.357 &       0.376 \\
PG0850$+$170 &  27.81 &        0.466 \\ \hline
\end{tabular}
\end{table}

\clearpage
\begin{figure*}
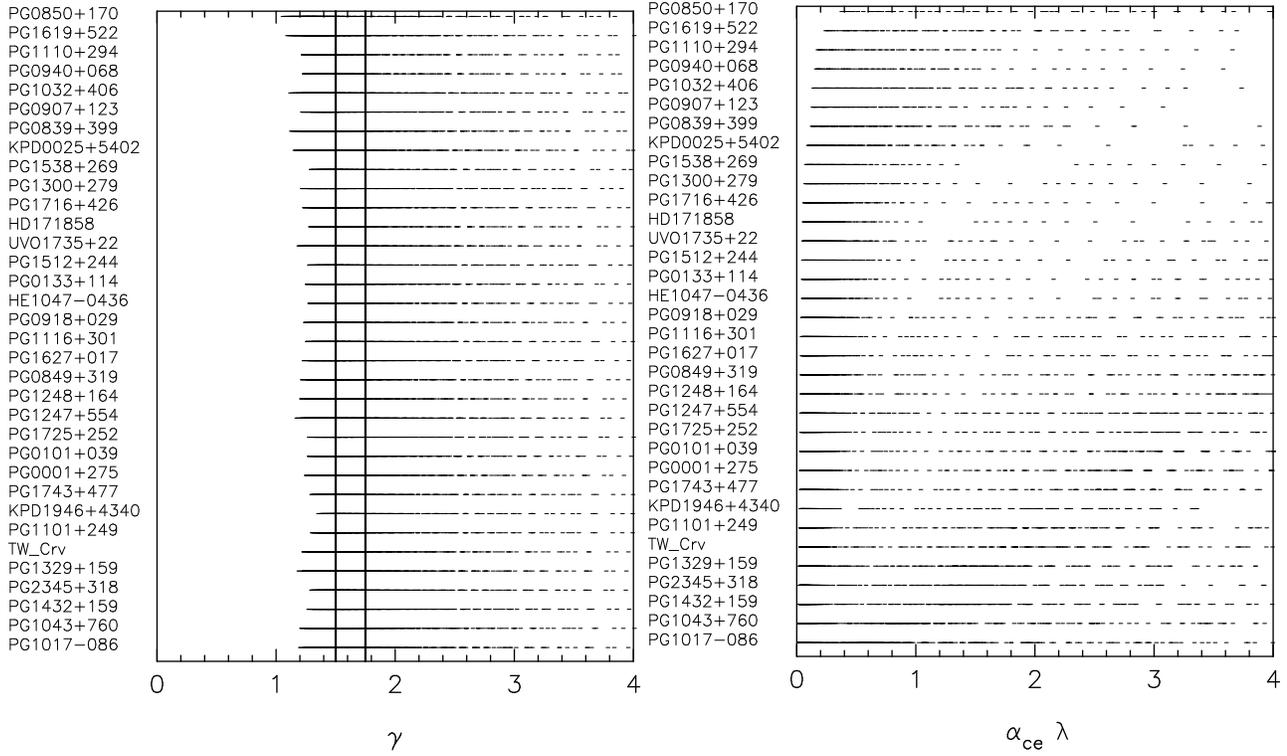

\resizebox{\columnwidth}{!}{\includegraphics[]{gamma_sdB1_1211.ps}}
\resizebox{\columnwidth}{!}{\includegraphics[]{al_sdB1_1211.ps}}
\caption[]{Left: Reconstructed $\gamma$ values for the mass transfer
  in sdB binaries in which the mass of the companion is unknown. From
  the radial velocities of the sdB star and an assumed sdB mass of 0.5
  \msun a minimum mass is inferred. For the upper limit a mass three
  times the minimum mass is assumed. Right: reconstructed $\alpha
  \lambda$ values for the same.}
\label{fig:gamma_al_sdB}
\end{figure*}

\end{document}